\documentstyle[preprint,prd,aps]{revtex}
\def\fmslash{\@ifnextchar[{\fmsl@sh}{\fmsl@sh[0mu]}}
\def\fmsl@sh[#1]#2{%
  \mathchoice
    {\@fmsl@sh\displaystyle{#1}{#2}}%
    {\@fmsl@sh\textstyle{#1}{#2}}%
    {\@fmsl@sh\scriptstyle{#1}{#2}}%
    {\@fmsl@sh\scriptscriptstyle{#1}{#2}}}
\def\@fmsl@sh#1#2#3{\m@th\ooalign{$\hfil#1\mkern#2/\hfil$\crcr$#1#3$}}
\makeatother
\begin{document}
\draft\pagenumbering{roma}
%%%%%%%%%%%%%%%%%%%%%%%%%%%%%%%%%%%%%%%%%%%%%%%%%%%%%%%%%%%%%%%%%%%%%%%%%%%%%%
\author{Ming-Qiu Huang}
\address{CCAST (World Laboratory) P.O. Box 8730, Beijing, 100080}
\address{and Department of Applied Physics, Changsha Institute of Technology,
Hunan 410073, China}
\author{Yuan-Ben Dai}
\address{Institute of Theoretical Physics, Academia Sinica,
P.O.Box 2735, Beijing 100080, China}
\title{Subleading Isgur-Wise form factors and ${\cal{O}}(1/m_Q)$ corrections to
the semileptonic decays $B\to D_1\ell\bar\nu$ and $B\to
D_2^*\ell\bar\nu$} \vspace{8mm}
\date{\today}
\maketitle
\thispagestyle{empty}
\vspace{15mm}

\begin{abstract}
 Exclusive semileptonic decays $B\to D_1(2420)\ell\bar\nu$ and
$B\to D_2^*(2460)\ell\bar\nu$ are studied at the subleading order of the heavy quark
expansion. The subleading Isgur-Wise functions resulted from the kinetic energy and
chromomagnetic corrections to the HQET Lagrangian are calculated by QCD sum rules in
the framework of the heavy quark effective theory. The decay rates and branching
ratios are computed with the inclusion of the order of $1/m_Q$ corrections. It is
found that the $1/m_Q$ correction to the decay rate is not large for $B \to D^*_2$
but is very large for $B\to D_1$.
\end{abstract}
\vspace{4mm}

\vspace{3.cm} %\noindent April 1998
\newpage
\pagenumbering{arabic}
%%%%%%%%%%%%%%%%%%%%%%%%%%%%%%%%%%%%%%%%%%%%%%%%%%%%%%%%%%%%%%%%%%%%%%%%%%%%%%%%

\section{Introduction}
\label{sec1}

In recent years there has been a continuous interest in the investigation of
semileptonic decays of $B$ meson into excited charmed mesons. This interest arises
from several reasons. The current experimental data show that the exclusive $B$
transitions to the ground state $s$-wave $D$ and $D^*$ mesons make up only
approximately 60\% of the inclusive semileptonic decay rate, thus a sizeable part of
semileptonic $B$ decays should go to excited $D$ meson states. Indeed, these decays
have been observed and more experimental data are collected with an increasing
accuracy \cite{cleo,aleph,opal,delphi}. Theoretically, the semileptonic $B$ decays
into excited charmed meson states can provide an additional source of information
for determining the CKM matrix element $V_{cb}$ as well as exploring the internal
dynamics of systems containing heavy-light quarks.

The heavy quark symmetry \cite{HQET,neubert1} has important consequences on the
spectroscopy and weak decay matrix elements of mesons containing a single heavy
quark $Q$. In the infinite mass limit, the spin and parity of the heavy quark and
that of the light degrees of freedom are separately conserved. This allows that the
hadronic states can be classified  in degenerate doublets by the total angular
momentum $j$ and the angular momentum of the light degrees of freedom $j_\ell$. In
the case of $\bar q Q$ mesons, coupling  $j_\ell$ with the spin of heavy quark
$s_Q=1/2$ yields  a doublet  with total spin $j=j_\ell\pm 1/2$. The ground state
mesons with $j_\ell^{P}=\frac12^-$ are the doublet ($D$,$D^*$) for $Q=c$ and
($B$,$B^*$) for $Q=b$. The excited heavy mesons with $j_\ell^P=1/2^+$ and $3/2^+$
can be classified in two doublets of spin symmetry ($0^+$,$1^+$) and ($1^+$,$2^+$),
which are identified as ($D'_0$, $D'_1$) and ($D_1$, $D_2^*$) for charmed mesons,
respectively. The other important application of heavy quark symmetries has been the
study of semileptonic transitions between two heavy hadrons. The hadronic matrix
elements of weak currents between members of the doublets identified by $j_\ell$ and
$j_{\ell'}$ can be expressed in terms of universal form factors which are functions
of the dot-product, $y=v\cdot v'$, of the initial and final hadron four-velocities.
A well-known result is that the semileptonic $B$ decays to ground state $D^{(*)}$
mesons, in the $m_Q\to\infty$ limit, can be described in terms of a single universal
function, the Isgur-Wise function $\xi(y)$. While the decays to $P$-wave excited $D$
mesons with $j_\ell^P=1/2^+$ and $3/2^+$ require two independent functions,
$\zeta(y)$ and $\tau(y)$ \cite{IWsr,Leib}, respectively, in the limit
$m_Q\to\infty$.

There are  $\Lambda_{\mathrm QCD}/m_Q$ corrections to the weak matrix elements
parameterized by form factors at the $m_Q\to\infty$ limit. The $\Lambda_{\mathrm
QCD}/m_Q$  corrections to the leading term can be analyzed in a systematic way in
heavy quark effective theory (HQET) in terms of a reduced number of universal
parameters. The $\Lambda_{QCD}/m_Q$ corrections may play an important role for
$B$-decay modes into excited charmed states since the corresponding transition
matrix elements in the infinite mass limit vanish at zero recoil point because of
heavy quark spin symmetry, while $\Lambda_{QCD}/m_Q$ corrections to these decay
matrix elements can give nonzero contributions at zero recoil \cite{Leib}. The
kinematically allowed range for these decays mostly occurs near the zero recoil
point, thus the magnitude of $\Lambda_{QCD}/m_Q$ corrections might be comparable
with the leading order result.

The universal functions must be estimated in some nonperturbative approaches. A
viable approach is the QCD sum rules \cite{svzsum} formulated in the framework of
HQET \cite{hqetsum}. This method allows us to relate hadronic observables to QCD
parameters {\it via} the operator product expansion (OPE) of the correlator. A
fruitful application of QCD sum rules has been the determination of the Isgur-Wise
functions parameterizing the $B\to D^{(*)}$ semileptonic transitions up to the
$\Lambda_{\mathrm QCD}/m_Q$ corrections\cite{BBBG,shifman,BBBG1,neubert}. The QCD
sum rule analysis for the semileptonic $B$ decays to excited $D$ mesons involves the
determination of the universal form factors. At leading order in the $1/m_Q$
expansion, the two independent universal form factors $\zeta(y)$ and $\tau(y)$, that
parameterize transitions $B\to D^{**}$ ($D^{**}$ being the generic $L=1$ charmed
state), have been calculated with QCD sum rules\cite{c-sum,h-dai}. Moreover,
perturbative corrections to ${\cal O}(\alpha_s)$ have been included in the QCD sum
rule for $\zeta(y)$ in \cite{cnew}. The other approaches include various versions of
the constituent quark model \cite{godfrey,iw2,cccn,wambach,veseli,oliver,DDG} and
relativistic Bethe-Salpeter equations \cite{dai2}. The analysis of $\Lambda_{\mathrm
QCD}/m_Q$ corrections is an important issue for the semileptonic $B$ decays to
excited $D$ mesons. Such corrections have been investigated in terms of meson mass
splittings in Ref. \cite{Leib} and by employing the relativistic quark model in Ref.
\cite{EFG}. The corrections have also been included by a variant approach in HQEFT
in Ref. \cite{w-wu}. At the order $1/m_Q$, the corrections for matrix elements of
$B\to D^{**}$ include contributions from higher-dimensional operators in the
effective currents and in the effective Lagrangian. For the semileptonic transitions
$B\to D_1\ell\bar\nu$ and $B\to D^*_2\ell\bar\nu$, the former give rise to two
independent universal functions, denoted by $\tau_1(y)$ and $\tau_2(y)$\cite{Leib}.
In the framework of QCD sum rules, these two independent form factors have been
investigated in our previous work in \cite{hldai}. Here we shall focus on the second
type of corrections, which originate from higher-order HQET effective Lagrangian.

The remainder of this paper is organized as follows. In Sec. \ref{sec2}  we review
the formulas for the matrix elements of the weak currents including the structure of
the $\Lambda_{\mathrm QCD}/m_Q$ corrections in the heavy quark effective theory. The
QCD sum rule analysis for the subleading Isgur-Wise functions related to the
corrections from the insertions of the kinetic energy and chromomagnetic operators
is presented in Sec. \ref{sec3}. Section \ref{sec4} is devoted to numerical results.
Concluding remarks are given in Sec. \ref{conc}.

%%%%%%%%%%%%%%%%%%%%%%%%%%%%%%%%%%%%%%%%%%%%%%%%%%%%%%%%%%%%%%%%%%%%%%%%%%%%%
\section{ The heavy-quark expansion and the subleading Isgur-Wise form factors}
\label{sec2}

The theoretical description of semileptonic decays involves the matrix
elements of vector and axial vector currents
 ($V^\mu=\bar c\,\gamma^\mu\,b$ and $A^\mu=\bar c\,\gamma^\mu\gamma_5\,b$)
between $B$ mesons and excited $D$ mesons. For the processes $B\to
D_1\ell\bar\nu$ and $B\to D_2^*\ell\bar\nu$, these matrix elements can be
parameterized as
\begin{mathletters}\label{matrix1}
\begin{eqnarray}%\label{matrix1}
{\langle D_1(v',\epsilon)|\,V^\mu\,|B(v)\rangle}
  &=&f_{V_1}\epsilon^{*\mu}+(f_{V_2}v^\mu+f_{V_3}v'^\mu)\,
  \epsilon^*\cdot v \,,\label{d1-v}\\*
{\langle D_1(v'
,\epsilon)|\,A^\mu\,|B(v)\rangle}
  &=& i\,f_A\, \varepsilon^{\mu\alpha\beta\gamma}
  \epsilon^*_\alpha v_\beta v'_\gamma \,,  \label{d1-a}\\*
{\langle D^*_2(v',\epsilon)|\, A^\mu\, |B(v)\rangle}
  &=&k_{A_1}\, \epsilon^{*\mu\alpha} v_\alpha
  + (k_{A_2} v^\mu + k_{A_3} v'^\mu)\,
  \epsilon^*_{\alpha\beta}\, v^\alpha v^\beta \,, \label{d2-a} \\*
{\langle D^*_2(v',\epsilon)|\, V^\mu\, |B(v)\rangle}
  &=& i\;k_V\, \varepsilon^{\mu\alpha\beta\gamma}
  \epsilon^*_{\alpha\sigma} v^\sigma v_\beta v'_\gamma \,.  \label{d2-v}
\end{eqnarray}\end{mathletters}
Here form factors $f_i$ and $k_i$ are dimensionless functions of $y$. In the above
equations we have used the mass-independent normalization $\langle M(v')| M(v)
\rangle = (2\pi)^3 {2p^0/m_M} \delta^3 (p-p')$ for the heavy meson states of
momentum $p=m_Mv $. Therefore, there is a different factor from the corresponding
equations in Ref. [8]. The differential decay rates expressed in terms of the form
factors are given by (taking the mass of the final lepton to zero)
\begin{eqnarray}
   \frac{{\mathrm d}\Gamma_{D_1}}{{\mathrm d}y}
   &=& \frac{G_F^2\,|V_{cb}|^2 m_B^5}{48\pi^3}\,r_1^3 \,\sqrt{y^2-1}
     \bigg\{ 2(1-2yr_1+r_1^2)\,
  \Big[ f_{V_1}^2 + (y^2-1)\,f_A^2 \Big] \nonumber\\*
&& + \Big[ (y-r_1)\,f_{V_1}
  + (y^2-1)\, (f_{V_3}+r_1 f_{V_2}) \Big]^2 \bigg\} \,, \label{drate1}\\
   \frac{{\mathrm d}\Gamma_{D^*_2}}
    {{\mathrm d}y}
   &=& \frac{G_F^2\,|V_{cb}|^2 m_B^5}{144\pi^3}\,r_2^3\,(y^2-1)^{3/2}
   \bigg\{ 3(1-2yr_2+r_2^2)\, \Big[ k_{A_1}^2 + (y^2-1)\,k_V^2 \Big] \nonumber\\*
&& + 2\Big[ (y-r_2)\,k_{A_1}
  + (y^2-1)\, (k_{A_3}+r_2\,k_{A_2}) \Big]^2 \bigg\} \,.\label{drate2}
\end{eqnarray}
where $r_1=m_{D_1}/m_B$ and $r_2=m_{D_2^*}/m_B$.

The form factors $f_i$ and $k_i$ can be expressed by a set of Isgur-Wise
functions at each order in $\Lambda_{\rm QCD}/m_{c,b}$. This is achieved by
evaluating the matrix elements of the effective current operators arising
from the HQET expansion of the weak currents. A convenient way to evaluate
hadronic matrix elements is by using the trace formalism developed in Ref.
\cite{Falk} to parameterize the matrix elements in Eqs. (\ref{matrix1}).
Following this method, one introduces the matrix representations
\begin{mathletters}\label{c-rep}
\begin{eqnarray}\label{c-rep1}
&&H_v = \frac{1+\fmslash v}2\, \Big[ P_v^{*\mu} \gamma_\mu
  - P_v\, \gamma_5 \Big] \,,\\
&&F_v^\mu = \frac{1+\fmslash v}2 \bigg\{ \! P_v^{*\mu\nu} \gamma_\nu
  - \sqrt{\frac32}\, P_v^\nu \gamma_5 \bigg[ g^\mu_\nu -
  \frac13 \gamma_\nu (\gamma^\mu-v^\mu) \bigg] \bigg\} \;,  \label{c-rep2}
\end{eqnarray}\end{mathletters}
where $P_v$, $P_v^{*\mu}$ and $P_v^\nu$, $P_v^{*\mu\nu}$ are annihilation
operators for members of the $j_\ell^P=1/2^-$ and $3/2^+$ doublets with
four-velocity $v$ in HQET. The matrices $H$ and $F$ satisfy $\fmslash v
H_v=H_v=-H_v\fmslash v$,~ $\fmslash v F_v^\mu=F_v^\mu=-F_v^\mu\fmslash v$,~
$F_v^\mu\gamma_\mu=0$, and $v_\mu F_v^\mu=0$.

At the leading order of the heavy quark expansion the hadronic matrix elements
of weak current between the states annihilated by the fields in $H_v$ and
$F_{v'}^\sigma$ are written as
\begin{eqnarray}
\label{leading}
\bar h^{(c)}_{v'}\, \Gamma\, h^{(b)}_v = \tau\;
  {\rm Tr}\, \Big\{ v_\sigma \bar F^\sigma_{v'}\, \Gamma\, H_v \Big\} \,.
\end{eqnarray}
where $h_v^{(Q)}$ is the heavy quark field in the effective theory and
 $\tau$ is a universal Isgur-Wise function of $y$.

At the order $\Lambda_{\rm QCD}/m_Q$ there are contributions to the decay matrix
elements originating from corrections to the HQET Lagrangian of
the same order
 \begin{eqnarray}\label{dlagr}
\delta{\cal L}=\frac{1}{2 m_Q} \Big[ O_{{\rm kin},v}^{(Q)} +
  O_{{\rm mag},v}^{(Q)} \Big]\;,
 \end{eqnarray}
 where
 \begin{eqnarray}
  O_{{\rm kin},v}^{(Q)} = \bar h_v^{(Q)} (iD)^2 h_v^{(Q)}, \hspace {1.2cm}
  O_{{\rm mag},v}^{(Q)} = \bar h_v^{(Q)}
  \frac{g_s}2 \sigma_{\alpha\beta} G^{\alpha\beta} h_v^{(Q)}\;.\nonumber
\end{eqnarray}
The matrix elements of $\Lambda_{\rm QCD}/m_Q$ corrections  from
the insertions of the kinetic energy operator $O_{\rm kin}$ and
chromomagnetic operator $O_{\rm mag}$ can be parameterized as
\begin{eqnarray}\label{corr-lag}
i \int {\rm d}^4x\, T\,\Big\{ O_{{\rm kin},v'}^{(c)}(x)\,
  \Big[ \bar h_{v'}^{(c)}\, \Gamma\, h_{v}^{(b)} \Big](0)\, \Big\}
  &=& \eta^{(c)}_{\rm ke}\, {\rm Tr}\, \Big\{ v_\sigma
  \bar F^\sigma_{v'}\, \Gamma\, H_v \Big\} \,, \\
i \int {\rm d}^4x\, T\,\Big\{ O_{{\rm kin},v}^{(b)}(x)\,
  \Big[ \bar h_{v'}^{(c)}\, \Gamma\, h_{v}^{(b)} \Big](0)\, \Big\}
  &=& \eta^{(b)}_{\rm ke}\, {\rm Tr}\, \Big\{ v_\sigma
  \bar F^\sigma_{v'}\, \Gamma\, H_v \Big\} \,;\\
i \int {\rm d}^4x\, T\,\Big\{ O_{{\rm mag},v'}^{(c)}(x)\,
  \Big[ \bar h_{v'}^{(c)}\, \Gamma\, h_{v}^{(b)} \Big](0)\, \Big\}
  &=& {\rm Tr}\, \bigg\{ {\cal R}_{\sigma\alpha\beta}^{(c)}\,
  \bar F_{v'}^\sigma\, i\sigma^{\alpha\beta}\, \frac{1+\fmslash{v'}}2\,
  \Gamma\, H_v \bigg\} \,, \\*
i \int {\rm d}^4x\, T\,\Big\{ O_{{\rm mag},v}^{(b)}(x)\,
  \Big[ \bar h_{v'}^{(c)}\, \Gamma\, h_{v}^{(b)} \Big](0)\, \Big\}
  &=& {\rm Tr}\, \bigg\{ {\cal R}_{\sigma\alpha\beta}^{(b)}\,
  \bar F_{v'}^\sigma\, \Gamma\, \frac{1+\fmslash{v}}2\, i\sigma^{\alpha\beta}
  H_v \bigg\} \,.
\end{eqnarray}
The functions $\eta^{(c,b)}_{\rm ke}(y)$ have mass dimension and effectively
correct the leading order Isgur-Wise function $\tau(y)$ since the kinetic
energy operator does not violate heavy quark spin symmetry. The most general
decomposition for ${\cal R}^{(c,b)}$ are
\begin{eqnarray}\label{Decom}
{\cal R}_{\sigma\alpha\beta}^{(c)} &=&
  \eta_1^{(c)}\, v_\sigma \gamma_\alpha \gamma_\beta
  + \eta_2^{(c)}\, v_\sigma v_\alpha \gamma_\beta
  + \eta_3^{(c)}\, g_{\sigma\alpha} v_\beta \,, \nonumber\\*
{\cal R}_{\sigma\alpha\beta}^{(b)} &=&
  \eta_1^{(b)}\, v_\sigma \gamma_\alpha \gamma_\beta
  + \eta_2^{(b)}\, v_\sigma v'_\alpha \gamma_\beta
  + \eta_3^{(b)}\, g_{\sigma\alpha} v'_\beta \,,
\end{eqnarray}
where $\eta_i$ are function of $y$, and have mass dimension one.

There are also order $\Lambda_{\rm QCD}/m_{c,b}$ corrections originating
from the matching of the $b\to c$ flavor changing current onto the effective
theory, they can be parameterized in terms of two independent Isgur-Wise
functions, $\tau_1$ and $\tau_2$ \cite{Leib}.

Summing up all the contributions up to order $\Lambda_{\rm QCD}/m_{c,b}$, it
is straightforward to express the form factors $f_i$ and $k_i$
parameterizing $B\to D_1\,\ell\,\bar\nu$ and $B\to D_2^*\,\ell\,\bar\nu_e$
semileptonic decays in terms of Isgur-Wise functions. The explicit
expressions for $f_i$ and $k_i$ are as follows \cite{Leib}:
\begin{eqnarray}\label{fkexp}
\sqrt6\, f_A &=& - (y+1)\tau
  - \varepsilon_b \{ (y-1) [(\bar\Lambda'+\bar\Lambda)\tau
  - (2y+1)\tau_1-\tau_2] + (y+1)\eta_b \} \nonumber\\*
&& - \varepsilon_c [ 4(y\bar\Lambda'-\bar\Lambda)\tau - 3(y-1) (\tau_1-\tau_2)
  + (y+1) (\eta_{\rm ke}-2\eta_1-3\eta_3) ] \,,\nonumber\\*
\sqrt6\, f_{V_1} &=&  (1-y^2)\tau
  - \varepsilon_b (y^2-1) [(\bar\Lambda'+\bar\Lambda)\tau
  - (2y+1)\tau_1-\tau_2 + \eta_b] \nonumber\\*
&& - \varepsilon_c [ 4(y+1)(y\bar\Lambda'-\bar\Lambda)\tau
  - (y^2-1)(3\tau_1-3\tau_2-\eta_{\rm ke}+2\eta_1+3\eta_3) ] \,, \nonumber\\
\sqrt6\, f_{V_2} &=& -3\tau - 3\varepsilon_b [(\bar\Lambda'+\bar\Lambda)\tau
  - (2y+1)\tau_1-\tau_2 + \eta_b] \nonumber\\*
&& - \varepsilon_c [ (4y-1)\tau_1+5\tau_2 +3\eta_{\rm ke} +10\eta_1
  + 4(y-1)\eta_2-5\eta_3 ] \,, \nonumber\\*
\sqrt6\, f_{V_3} &=&  (y-2)\tau
  + \varepsilon_b \{ (2+y) [(\bar\Lambda'+\bar\Lambda)\tau
  - (2y+1)\tau_1-\tau_2] - (2-y)\eta_b \} \nonumber\\*
&& + \varepsilon_c [ 4(y\bar\Lambda'-\bar\Lambda)\tau +
  (2+y)\tau_1 + (2+3y)\tau_2   + (y-2)\eta_{\rm ke} \nonumber\\*
&&- 2(6+y)\eta_1 - 4(y-1)\eta_2 - (3y-2)\eta_3 ] \,\\*
k_V &=& - \tau - \varepsilon_b [(\bar\Lambda'+\bar\Lambda)\tau
  - (2y+1)\tau_1-\tau_2 + \eta_b]
  - \varepsilon_c (\tau_1-\tau_2+\eta_{\rm ke}-2\eta_1+\eta_3) , \nonumber\\*
k_{A_1} &=& - (1+y)\tau - \varepsilon_b \{ (y-1)
  [(\bar\Lambda'+\bar\Lambda)\tau - (2y+1)\tau_1-\tau_2] + (1+y)\eta_b \}
  \nonumber\\*
&& - \varepsilon_c [ (y-1)(\tau_1-\tau_2)
  + (y+1)(\eta_{\rm ke}-2\eta_1+\eta_3) ] , \nonumber\\
k_{A_2} &=& - 2\varepsilon_c (\tau_1+\eta_2) , \nonumber\\*
k_{A_3} &=& \tau + \varepsilon_b [(\bar\Lambda'+\bar\Lambda)\tau
  - (2y+1)\tau_1-\tau_2 + \eta_b]
  - \varepsilon_c (\tau_1+\tau_2-\eta_{\rm ke}+2\eta_1-2\eta_2-\eta_3)\;,\nonumber
\end{eqnarray}
where $\varepsilon_Q=1/(2m_Q)$, $\eta_{ke} = \eta_{ke}^c$ and $\eta_b=\eta_{\rm ke}^{(b)}+6\,\eta_1^{(b)}-2(y-1)\,%
\eta_2^{(b)}+\eta_3^{(b)}$, $\bar\Lambda$($\bar\Lambda'$) is mass parameter of ground state
(excited) mesons in HQET and the superscript on $\tau_i^{(c)}$ and
$\eta_i^{(c)}$ are dropped.

The form factors $\tau$ and $\tau_i$ ($i=1,2$) in HQET, that occur in Eq.
(\ref{fkexp}) have been investigated by using QCD sum rules in our previous
work \cite{h-dai,hldai}. In the following sections we shall extend the QCD
sum rule analysis to the calculation of the subleading Isgur-Wise functions,
$\eta_{\rm ke}(y)$ and $\eta_i(y)$, associated with the insertions of
kinetic energy and chromomagnetic operators of the HQET Lagrangian, $\delta
{\cal L}$ in Eq.~(\ref{dlagr}).

%%%%%%%%%%%%%%%%%%%%%%%%%%%%%%%%%%%%%%%%%%%%%%%%%%%%%%%%%%%%%%%%%%%%%%%%%%%%%%%%
\section{QCD sum rules for Isgur-Wise functions $\eta_{ke}$ and $\eta_i$}
\label{sec3}
%%%%%%%%%%%%%%%%%%%%%%%%%%%%%%%%%%%%%%%%%%%%%%%%%%%%%%%%%%%%%%%%%%%%%%%%%%%%%%%

A basic element in the application of QCD sum rules to problems involving excited
heavy mesons is to choose a set of appropriate interpolating currents in terms of
quark fields each of which creates (annihilates) an excited state of the heavy meson
with definite quantum numbers $j$, $P$, $j_\ell$.  The proper interpolating current
$J_{j,P,j_{\ell}}^{\alpha_1\cdots\alpha_j}$ for the state with  arbitrary quantum
number $j$, $P$, $j_{\ell}$ in HQET was given in \cite{huang}. These currents have
nice properties. They were proven to satisfy the following conditions
\begin{eqnarray}
\label{decay}
\langle 0|J_{j,P,j_{\ell}}^{\alpha_1\cdots\alpha_j}(0)|j',P',j_{\ell}^{'}\rangle
&=&
f_{Pj_l}\delta_{jj'}
\delta_{PP'}\delta_{j_{\ell}j_{\ell}^{'}}\eta^{\alpha_1\cdots\alpha_j}\;,\\
\label{corr}
i\:\langle 0|T\left (J_{j,P,j_{\ell}}^{\alpha_1\cdots\alpha_j}(x)J_{j',P',j_{\ell}'}^{\dag
\beta_1\cdots\beta_{j'}}(0)\right )|0\rangle&=&\delta_{jj'}\delta_{PP'}\delta_{j_{\ell}j_{\ell}'}
(-1)^j\:{\cal S}\:g_t^{\alpha_1\beta_1}\cdots g_t^{\alpha_j\beta_j}\nonumber\\[2mm]&&\times\:
\int \,dt\delta(x-vt)\:\Pi_{P,j_{\ell}}(x)
\end{eqnarray}
in the $m_Q\to\infty$ limit, where $\eta^{\alpha_1\cdots\alpha_j}$ is the
polarization tensor for the spin $j$ state,  $v$ is the velocity of the heavy quark,
$g^{\alpha\beta}_t=g^{\alpha\beta}-v^\alpha v^\beta$ is the transverse metric
tensor, ${\cal S}$ denotes symmetrizing the indices and subtracting the trace terms
separately in the sets $(\alpha_1\cdots\alpha_j)$ and $(\beta_1\cdots\beta_{j})$,
$f_{P,j_{\ell}}$ and $\Pi_{P,j_{\ell}}$ are a constant and a function of $x$
respectively which depend only on $P$ and $j_{\ell}$. Because of Eqs. (\ref{decay})
and (\ref{corr}), the sum rules in HQET for decay amplitudes derived from a
correlator containing such currents receive a contribution only from one of the two
states with the same spin-parity $(j,P)$ but different $j_\ell$ in the
$m_Q\to\infty$. Starting from the calculations in the leading order, the decay
amplitudes for finite $m_Q$ can be calculated unambiguously order by order in the
$1/m_Q$ expansion in HQET.

Following \cite{huang} the local interpolating current for creating $0^-$
pseudoscalar $B$ meson is taken as
\begin{eqnarray}
\label{p-scalar} J^{\dag}_{0,-,{1/2}}=\sqrt{\frac{1}{2}}\:\bar
h_v\gamma_5q\;,
\end{eqnarray}
and the local interpolating currents for creating $1^+$ and $2^+$
($D_1$, $D_2^*$) mesons in the doublet ($D_1$, $D_2^*$) are taken as
\begin{eqnarray}
\label{curr-1}
J^{\dag\alpha}_{1,+,3/2}&=&\sqrt{\frac{3}{4}}\:\bar h_v\gamma^5(-i)\left(
{\cal D}_t^{\alpha}-\frac{1}{3}\gamma_t^{\alpha}\fmslash{\cal D}_t\right)q\;,\\
\label{curr-2}
J^{\dag\alpha_1,\alpha_2}_{2,+,3/2}&=&\sqrt{\frac{1}{2}}\:\bar h_v
\frac{(-i)}{2}\left(\gamma_t^{\alpha_1}{\cal D}_t^{\alpha_2}+
\gamma_t^{\alpha_2}{\cal D}_t^{\alpha_1}-\frac{2}{3}\;g_t^{\alpha_1\alpha_2}
\fmslash{\cal D}_t\right)q\;,
\end{eqnarray}
where ${\cal D}$ is the covariant derivative and
$\gamma_{t}^\mu=\gamma^\mu-\fmslash vv^\mu$. Note that, without
the last term in the bracket in (\ref{curr-1}) the current would
couple also to the $1^+$ state in the doublet $(0^+,1^+)$ even in
the limit of infinite $m_Q$.

The QCD sum rule calculations for the correlators of two heavy-light currents give:
\cite{neubert,huang}
\begin{eqnarray}
f_{-,{1/2}}^2\,e^{-2\bar\Lambda_{-,{1/2}}/T} &=&
\frac{3}{16\pi^2}\int_0^{\omega_{c1}}\omega^2e^{-\omega/{T}} d\omega -
\frac12\langle\bar q q\rangle\left(1- \frac{m_0^2}{4 T^2}\right)
\,,\label{2-point1}\\ f_{+,{3/2}}^2e^{-2\bar\Lambda_{+,{3/2}}/{T}}&=&{1\over
2^6\pi^2} \int_0^{\omega_{c2}}\omega^4e^{-\omega/{T}}d\omega
-\frac{1}{12}\:m_0^2\:\langle\bar qq\rangle-{1\over
2^5}\langle{\alpha_s\over\pi}G^2\rangle T\;,\label{2-point2}
\end{eqnarray}
where $m_0^2\,\langle\bar qq\rangle=\langle\bar
qg\sigma_{\mu\nu}G^{\mu\nu}q\rangle$.

The QCD sum rule analysis for the subleading form factors proceeds along the same
lines as that for the leading order Isgur-Wise function. For the determination of
the form factor $\eta_{\rm ke}$, which relates to the insertion of $\Lambda_{\rm
QCD}/m_c$ kinetic operator of the HQET Lagrangian, one studies the analytic
properties of the three-point correlators
\begin{mathletters}\label{3-point1}
\begin{eqnarray}
 i^2\int\, d^4xd^4x'd^4z\,e^{i(k'\cdot x'-k\cdot x)}\;\langle 0|T\left(
 J^{\nu}_{1,+,3/2}(x')\;O_{{\rm kin},v'}^{(c)}(z)\,{\cal J}^{\mu(v,v')}_{V,A}(0)\;
 J^{\dagger}_{0,-,1/2}(x)\right)|0\rangle \nonumber\\=
 \Xi(\omega,\omega',y)\;{\cal L}^{\mu\nu}_{V,A}\;, \hspace{3cm}\\
i^2\int\, d^4xd^4x'd^4z\,e^{i(k'\cdot x'-k\cdot x)}\;\langle
0|T\left(
 J^{\alpha\beta}_{2,+,3/2}(x')\;O_{{\rm kin},v'}^{(c)}(z)\,{\cal J}^{\mu(v,v')}_{V,A}(0)\;
 J^{\dagger}_{0,-,1/2}(x)\right)|0\rangle\nonumber\\=
 \Xi(\omega,\omega',y)\;{\cal L}^{\mu\alpha\beta}_{V,A}\;,\hspace{3cm}
\end{eqnarray}
\end{mathletters}
where ${\cal J}^{\mu(v,v')}_{V}=\bar h(v')\gamma^\mu\,h(v)$, ${\cal
J}^{\mu(v,v')}_{A}=\bar h(v')\gamma^\mu\gamma_5\,h(v)$. The variables $k$, $k'$
denote residual ``off-shell" momenta which are related to the momenta $P$ of the
heavy quark in the initial state and $P'$ in the final state by $k=P-m_Qv$,
$k'=P'-m_{Q'}v'$, respectively. For heavy quarks in bound states they are
typically of order $\Lambda_{QCD}$ and remain finite in the heavy quark limit.
${\cal L}_{V,A}$ are Lorentz structures  associated with the vector and axial
vector currents(see Appendix).

The coefficient $\Xi(\omega,\omega',y)$ in (\ref{3-point1}) is an analytic
function in the ``off-shell energies" $\omega=2v\cdot k$ and $\omega'=2v'\cdot
k'$ with discontinuities for positive values of these variables. It furthermore
depends on the velocity transfer $y=v\cdot v'$, which is fixed in its physical
region for the process under consideration. By saturating (\ref{3-point1}) with
physical intermediate states in HQET, one can isolate the contribution of
interest as the one having poles at $\omega=2\bar\Lambda_{-,{1/2}}$,
$\omega'=2\bar\Lambda_{+,{3/2}}$. Notice that the insertions of the kinetic
operator not only renormalize the leading Isgur-Wise function, but also the meson
coupling constants and the physical masses of the heavy mesons which define the
position of the poles. The correct hadronic representation of the correlator is
\begin{eqnarray}
\label{pole} \Xi_{\rm hadro}(\omega,\omega',y)&=&{f_{-,{1/2}}f_{+,{3/2}} \over
(2\bar\Lambda_{-,{1/2}}-\omega- i\epsilon )(2\bar\Lambda_{+,{3/2}}-\omega'-
i\epsilon)}\;\bigg(\eta_{\rm ke}(y)+\nonumber\\&& (G_{+,3/2}^K+\frac{K_{+,3/2}}
{2\bar\Lambda_{+,{3/2}}-\omega'- i\epsilon})\;\tau(y)\bigg)+{\rm higher
~resonance} \;,
\end{eqnarray}
where $f_{P,j_\ell}$ are constants defined in (\ref{decay}),
$\bar\Lambda_{P,j_\ell}=m_{P,j_\ell}-m_Q$, and $K_{P,j_\ell}$ and
$G_{P,j_\ell}^K$ are defined by \cite{huang,dai-zhu}
\begin{mathletters}\label{G-K}
\begin{eqnarray}
 \langle j,P,j_\ell|O_{{\rm
kin},v}^{(Q)}|j,P,j_\ell\rangle&=&K_{P,j_\ell}\;,\\\ \langle
0|i\int d^4x\;O_{{\rm
kin},v}^{(Q)}(x)J_{j,P,j_\ell}^{\alpha_1\cdots\alpha_j}(0)
|j,P,j_\ell\rangle &=&f_{P,j_\ell}\;G_{P,j_\ell}^K
\eta^{\alpha_1\cdots\alpha_j}\;.
\end{eqnarray}\end{mathletters}
Furthermore, as the result of equation (\ref{decay}), only one state with
$j^P=1^+$ or $j^P=2^+$ contributes to (\ref{pole}), the other resonance with
the same quantum number $j^P$ and different $j_l$ does not contribute. This
would not be true for $j^P=1^+$ if the last term in (\ref {curr-1}) is
absent.

Following the standard QCD sum rule procedure the calculations of
$\Xi(\omega,\omega',y)$ are straightforward. In doing this, for
simplicity, the residual momentum $k$ is chosen to be parallel to
$v$ such that $k_\mu=(k\cdot v)v_\mu$ (and similar for $k'$).
Confining us to the leading order of perturbation and the
operators with dimension $D\leq 5$ in OPE, the relevant Feynman
diagrams are shown in Fig 1. The perturbative part of the spectral
density is
\begin{eqnarray}
\label{spectral1}
 \rho_{\rm pert}(\tilde\omega,\tilde\omega',y)&=&\frac{3}{2^8\pi^2}
 \frac{1}{(1+y)(y^2-1)^{5/2}}
 \bigg((2y-3)\tilde\omega^2+(8y^2+12y-1)\tilde\omega^{'2}\nonumber\\
&& \quad\mbox{}+(6y^2+18y-3)\tilde\omega^2\tilde\omega'
 -(12y^3+18y^2+9)\tilde\omega\tilde\omega^{'2}\bigg)\;\nonumber\\
&& \quad\mbox{}
\times\Theta(\tilde\omega)\,\Theta(\tilde\omega')\,
\Theta(2y\tilde\omega\tilde\omega'-\tilde\omega^2-\tilde\omega^{'2})\;.
\end{eqnarray}

The QCD sum rule is obtained by equating the phenomenological and
theoretical expressions for $\Xi$. In doing this the quark-hadron duality
needs to be assumed to model the contributions of higher resonance part of
Eq. (\ref{pole}). Generally speaking, the duality is to simulate the
resonance contribution by the perturbative part above some threshold
energies. In the QCD sum rule analysis for  $B$ semileptonic decays into
ground state $D$ mesons, it is argued by Neubert, Block and Shifman in
\cite{neubert1,shifman,neubert} that the perturbative and the hadronic
spectral densities can not be locally dual to each other, the necessary way
to restore duality  is to integrate the spectral densities over the
``off-diagonal'' variable $\tilde\omega_-=(\tilde\omega-\tilde\omega')/2$,
keeping the ``diagonal'' variable
$\tilde\omega_+=(\tilde\omega+\tilde\omega')/2$ fixed. It is in
$\tilde\omega_+$ that the quark-hadron duality is assumed for the integrated
spectral densities. We shall use the same prescription in the case of $B$
semileptonic decays into excited state $D$ mesons.

The $\Theta$ functions in (\ref{spectral1}) imply that in terms of
$\tilde\omega_+$ and $\tilde\omega_-$ the double discontinuities of the
correlator are confined to the region
 $-\sqrt{y^2-1}/(1+y)\;\tilde\omega_+\leq\tilde\omega_-\leq\sqrt{y^2-1}/(1+y)\;\tilde\omega_+$
and $\tilde\omega_+\geq 0$. According to our prescription an isosceles triangle
with the base $\tilde\omega_+ = \tilde\omega_c$ is retained in the integration
domain of the perturbative term in the sum rule.

In view of the asymmetry of the problem at hand with respect to the initial
and final states one may attempt to use an asymmetric triangle in the
perturbative integral. However, in that case the factor $(y^2-1)^{5/2}$ in
the denominator of (\ref {spectral1}) is not canceled after the integration
so that the Isgur-Wise function or it's derivative will be divergent at
$y=1$. Similar situation occurs for the sum rule of the Isgur-Wise function
for transition between ground states if a different domain is taken in the
perturbative integral \cite{neubert}.

In order to suppress the contributions of higher resonance states a double
Borel transformation in $\omega$ and $\omega'$ is performed to both sides of
the sum rule, which introduces two Borel parameters $T_1$ and $T_2$. For
simplicity we shall take the two Borel parameters equal: $T_1 = T_2 =2T$. In
the following section we shall estimate the changes in the sum rules in the
case of $T_1 \neq T_2$.

The non-perturbative power corrections to the correlators are computed from
the diagrams involving the quark and gluon condensates in Fig. 1(b)-1(k) in
the Fock-Schwinger gauge $x_\mu A^\mu(x)=0$. We find that the only
nonvanishing contribution is the gluon  condensate. Note, in particular, the
vanishing of the mixed quark-gluon condensate ($D=5$) resulting from the
explicit calculation of the diagram shown in Fig. 1(b). After adding the
non-perturbative part and making the double Borel transformation one obtains
the sum rule for $\eta_{\rm ke}$ as follows
\begin{eqnarray}
\label{sum-rule1} \eta_{\rm ke}(y)\,f_{-,1/2}f_{+,3/2}\;e^{-(\bar\Lambda_{-,1/2}
+\bar\Lambda_{+,3/2})/T}=-(G_{+,3/2}^K+\frac{K_{+,3/2}}{2T})\;
\tau(y)\;f_{-,1/2}f_{+,3/2}\times\nonumber\\ e^{-(\bar\Lambda_{-,1/2}
+\bar\Lambda_{+,3/2})/T}-\frac{3}{8\pi^2}\frac{y+2}{(y+1)^4}\;\int_0^{\omega_c}d{\omega_+}\,
\omega_+^4\,e^{-\omega_+/T}+\frac{1}{96}\langle\frac{\alpha_s}{\pi}GG\rangle
\frac{15y-1}{(y+1)^3}T\;,
\end{eqnarray}
in which the sum rule for $\tau(y)$ has been derived from the study of the
three-point correlator in Ref. \cite{h-dai} as
\begin{eqnarray}
\label{sumrule-tau} \tau(y)\,f_{-,1/2}\,f_{+,3/2}\;e^{-(\bar\Lambda_{-,1/2}
+\bar\Lambda_{+,3/2})/T}&=&
\frac{1}{2\pi^2}\frac{1}{(y+1)^3}\;\int_0^{\omega_c}d{\omega_+}\,
\omega_+^3\,e^{-\omega_+/T}-\frac{1}{12}m^2_0{\langle\bar qq\rangle\over T}
\nonumber\\ && \quad\mbox{}-\frac{1}{3\times 2^5} \langle
\frac{\alpha_s}{\pi}GG\rangle\frac{y+5}{(y+1)^2}\;.
\end{eqnarray}

From the consideration of symmetry, the sum rule for $\eta_{\rm ke}^{(b)}$ that
originates from the insertion of $\Lambda_{\rm QCD}/m_b$ kinetic operator of the
HQET Lagrangian is of the same form as in (\ref{sum-rule1}), but with the HQET
parameters $G_{+,3/2}^{K}$ and $K_{+,3/2}$ replaced by $G_{-,1/2}^{K}$ and
$K_{-,1/2}$, respectively. The definitions of $G_{-,1/2}^{K}$ and $K_{-,1/2}$ can be
found in Eq. (\ref{G-K}).

It is worth noting that the QCD $O(\alpha_s)$ corrections have not been included in
the sum rule calculations. However, the Isgur-Wise function obtained from the QCD
sum rule actually is a ratio of the three-point correlator to the two-point
correlator results.  While both of these correlators are subject to large
perturbative QCD corrections, it is expected that their ratio is not much affected
by these corrections  because of cancellation. This has been  proved to be true in
the analysis of Ref. \cite{cnew}.

We now turn to the QCD sum rule calculations of the functions parameterizing  the
time-ordered products of the chromomagnetic term in the HQET Lagrangian with the
leading order currents, $\eta_i$ ($i=1,2,3$). To obtain QCD sum rules for these
universal functions one starts from three-point correlators
\begin{mathletters}\label{3-point2}
\begin{eqnarray}
 i^2\int\, d^4xd^4x'd^4z\,e^{i(k'\cdot x'-k\cdot x)}\;\langle 0|T\left(
 J^{\nu}_{1,+,3/2}(x')\;O_{{\rm mag},v'}^{(c)}(z)\,{\cal J}^{\mu(v,v')}_{V,A}(0)\;
 J^{\dagger}_{0,-,1/2}(x)\right)|0\rangle \nonumber\\=
 \Xi^{\mu\nu}_{V,A}(\omega,\omega',y)\;,\label{mag-d1} \hspace{2.8cm}\\
i^2\int\,d^4xd^4x'd^4z\,e^{i(k'\cdot x'-k\cdot x)}\;\langle
0|T\left(
 J^{\alpha\beta}_{2,+,3/2}(x')\;O_{{\rm mag},v'}^{(c)}(z)\,{\cal J}^{\mu(v,v')}_{V,A}(0)\;
 J^{\dagger}_{0,-,1/2}(x)\right)|0\rangle\nonumber\\=
 \Xi^{\mu\alpha\beta}_{V,A}(\omega,\omega',y)\;.\label{mag-d2}\hspace{2.8cm}
\end{eqnarray}
\end{mathletters}

By saturating the double dispersion integral for the three-point functions in
(\ref{3-point2}) with hadronic states, one can isolate the contributions from the
double pole at $\omega=2\bar\Lambda_{-,{1/2}}$, $\omega'=2\bar\Lambda_{+,{3/2}}$.
Similarly, the insertions of the chromomagnetic operator result in the
corrections to the leading Isgur-Wise function as well as to the couplings of the
heavy mesons to the interpolating currents and to the physical meson masses. It
follows from Eq. (\ref{3-point2}) that
\begin{eqnarray}
\Xi_V^{\mu\nu\rm pole}(\omega,\omega',y)&=&{f_{-,{1/2}}f_{+,{3/2}} \over
(2\bar\Lambda_{-,{1/2}}-\omega- i\epsilon)(2\bar\Lambda_{+,{3/2}}-\omega'-
i\epsilon)}\bigg(-\xi_{1}{\cal L}^{\mu\nu}_{V}+\nonumber\\&&\xi_{2}{\cal
L}^{\mu\nu}_{V\xi} +(G_{+,3/2}^{\Sigma}+\frac{\Sigma_{+,3/2}}
{2\bar\Lambda_{+,{3/2}}-\omega'- i\epsilon})\;d_{1,3/2}\tau(y)\;{\cal
L}^{\mu\nu}_{V} \bigg)\;,
\\ \Xi_A^{\mu\nu\rm
pole}(\omega,\omega',y)&=&{f_{-,{1/2}}f_{+,{3/2}} \over
(2\bar\Lambda_{-,{1/2}}-\omega- i\epsilon)(2\bar\Lambda_{+,{3/2}}-\omega'-
i\epsilon)}\bigg(-\xi_{1}+\nonumber\\&&(G_{+,3/2}^{\Sigma}+\frac{\Sigma_{+,3/2}}
{2\bar\Lambda_{+,{3/2}}-\omega'- i\epsilon})\,d_{1,3/2}\tau(y)\bigg){\cal
L}^{\mu\nu}_{A}\;,\\ \Xi_V^{\mu\alpha\beta\rm
pole}(\omega,\omega',y)&=&{f_{-,{1/2}}f_{+,{3/2}} \over
(2\bar\Lambda_{-,{1/2}}-\omega- i\epsilon)(2\bar\Lambda_{+,{3/2}}-\omega'-
i\epsilon)}\bigg(-\zeta+\nonumber\\&& +(G_{+,3/2}^{\Sigma}+\frac{\Sigma_{+,3/2}}
{2\bar\Lambda_{+,{3/2}}-\omega'- i\epsilon})\;d_{2,3/2}\tau(y) \bigg)\;{\cal
L}^{\mu\alpha\beta}_{V}\;,\\ \Xi_A^{\mu\alpha\beta\rm
pole}(\omega,\omega',y)&=&{f_{-,{1/2}}f_{+,{3/2}} \over
(2\bar\Lambda_{-,{1/2}}-\omega- i\epsilon)(2\bar\Lambda_{+,{3/2}}-\omega'-
i\epsilon)}\bigg(-\zeta\;{\cal L}^{\mu\alpha\beta}_{A}+\eta_2{\cal
L}^{\mu\alpha\beta}_ {A\eta_2}\nonumber\\&&(G_{+,3/2}^{\Sigma}+\frac{\Sigma_{+,3/2}}
{2\bar\Lambda_{+,{3/2}}-\omega'- i\epsilon})\,d_{2,3/2}\tau(y)\;{\cal
L}^{\mu\alpha\beta}_{A}\bigg)\;,
\end{eqnarray}
where $\xi_{1}=2\eta_{1}+3\eta_{3}$,
$\xi_{2}=-16\eta_{1}-4(y-1)\eta_{2}-4\eta_3$, $\zeta=2\eta_1-\eta_3$, the
quantities $\Sigma_{P,j_\ell}^{(c)}$ and
$G_{\Sigma,P,j_\ell}^{(Q)}$ are defined by
\cite{huang,dai-zhu}
\begin{mathletters}\label{G-Sigma}
\begin{eqnarray}
\langle j,P,j_\ell|O_{{\rm
mag},v}^{(Q)}|j,P,j_\ell\rangle&=&d_m\Sigma_{P,j_\ell}\;,\\
\displaystyle \langle 0|i\int d^4x\;O_{{\rm
mag},v}^{(Q)}(x)J_{j,P,j_\ell}^{\alpha_1\cdots\alpha_j}(0)
|j,P,j_\ell\rangle &=&d_mf_{P,j_\ell}\;G_{P,j_\ell}^{\Sigma}
\eta^{\alpha_1\cdots\alpha_j}\;,\\
d_m=d_{j,j_\ell},~~d_{j_\ell-1/2,j_\ell}=2j_\ell+2,&&d_{j_\ell+1/2,j_\ell}=-2j_\ell\;,\nonumber
\end{eqnarray}\end{mathletters}
${\cal L}^{\mu\nu}_{V }$, ${\cal L}^{\mu\nu}_{V\xi}$, ${\cal
L}^{\mu\alpha\beta}_{V}$, ${\cal L}^{\mu\nu}_{A}$,  ${\cal L}^{\mu\alpha\beta}_{A}$
and ${\cal L}^{\mu\alpha\beta}_{A\eta_2}$ are defined in Appendix \ref{appendix}.
The three-point correlators (\ref{3-point2}) can be expressed in QCD in terms of a
perturbative part and nonperturbative contributions, which are related to the
theoretical calculation in HQET. When we do not consider radiative corrections, the
insertions of the chromomagnetic operator only contribute to diagrams involving
gluon condensates and do not contribute to the perturbative diagrams since there is
no way to contract the gluon contained in $O_{\rm mag}$. That is, the leading
nonperturbative contributions are proportional to the gluon condensates, while the
leading perturbative contributions are of order $\alpha_s$ and come from the
two-loop radiative corrections to the quark loop. In general, the calculation of the
two-loop diagrams is rather cumbersome. In this article we shall neglect the
perturbative term of order $\alpha_s$ and only include the nonperturbative gluon
condensates without radiative corrections in the sum rules.

Within this approximation one can perform the calculation conveniently by
using the Fock-Schwinger gauge. After making the double Borel transformation,
the sum rules for $\eta_i(y)$ are obtained as follows
\begin{mathletters}\label{sum-eta}
\begin{eqnarray}\label{sum-rule2}
f_{-,1/2}\,f_{+,3/2}\eta_1(y)\;e^{-(\bar\Lambda_{-,1/2}+\bar\Lambda_{+,3/2})/T}&=&
-\frac{1}{2}(G^\Sigma_{+,3/2}+\frac{\Sigma_{+,3/2}}{2T})\tau(y)\;f_{-,1/2}\,f_{+,3/2}e^{-(\bar\Lambda_{-,1/2}+
\bar\Lambda_{+,3/2})/T}\nonumber\\&&+\frac{1}{480}\langle
\frac{\alpha_s}{\pi}GG\rangle\frac{3y+2}{(y+1)^2}\;{T}\; \;,\\
\eta_2(y)&=&0\;,\label{sum-rule3}\\
f_{-,1/2}\,f_{+,3/2}\;\eta_3(y)\;e^{-(\bar\Lambda_{-,1/2}+\bar\Lambda_{+,3/2})/T}&=
&2(G_{+,3/2}^{\Sigma}+\frac{\Sigma_{+,3/2}}{2T})\tau(y)\;f_{-,1/2}\,f_{+,3/2}e^{-(\bar\Lambda_{-,1/2}+
\bar\Lambda_{+,3/2})/T}\nonumber\\&&-\frac{1}{240}\langle
\frac{\alpha_s}{\pi}GG\rangle\frac{14+y}{(y+1)^2}\;{T}\; \;.\label{sum-rule4}
\end{eqnarray}
\end{mathletters}
 $\eta_{1,2,3}$ are expected to be small compared to $\Lambda_{\rm QCD}$
since the mass splitting between $D^*_2$ and $D_1$ is very small. This is
supported by the fact that the QCD sum rule calculations indicate that the
analogous functions parameterizing the contributons of the chromomagnetic
operator for $B\to D^{(*)}\,e\,\bar\nu_e$ decays are small \cite{neubert}. The
$\Lambda_{\mathrm QCD}/m_b$ correction associated with the insertion of
chromomagnetic operator of the HQET Lagrangian can be investigated in a similar
way.

%%%%%%%%%%%%%%%%%%%%%%%%%%%%%%%%%%%%%%%%%%%%%%%%%%%%%%%%%%%%%%%%%%%%%%%%%%%%%%%
\section{Numerical results and implications for $B$ decays}
\label{sec4}

 We now turn to the numerical evaluation of these sum rules and the phenomenological
 implications. For the QCD parameters entering the theoretical expressions, we take
 the standard values \cite{svzsum,hqetsum}
 \begin{eqnarray}
   \langle\bar q q\rangle &=& -(0.23\pm 0.02)^3~\mbox{GeV}^3
    \,, \nonumber\\
   \langle{\alpha_s\over\pi} GG\rangle &=& (0.012\pm0.004)~\mbox{GeV}^4 \,,
    \nonumber\\
   m_0^2 &=& (0.8\pm0.2)~\mbox{GeV}^2 \,.
\label{cond}
\end{eqnarray}
In order to obtain numerical results for $\eta_{\rm ke}(y)$, $\eta_{\rm ke}^b(y)$
and $\eta_i(y)$ ($i=1,2,3$) from the sum rules which are independent of specific
input values of $f$'s, $\bar\Lambda$'s and $\tau$, we adopt the strategy to evaluate
the sum rules by eliminating the explicit dependence on these quantities by using
the sum rules for them. Substituting the sum rules (\ref{2-point1}) and
(\ref{2-point2}) into the left side and the sum rule (\ref{sumrule-tau}) into the
right side of the sum rules (\ref{sum-rule1}) and (\ref{sum-eta}) for the
three-point correlators, we obtain expressions for the $\eta_{\rm ke}$, $\eta_{\rm
ke}^b$ and $\eta_i$ ($i=1,2,3$) as functions of the Borel parameter $T$ and the
continuum thresholds. This procedure may help to reduce the uncertainties in the
calculation. For other HQET parameters we use the following values obtained by QCD
sum rules \cite{huang,hhh,dai-zhu,bball}:
\begin{eqnarray}
&&K_{+,3/2}=-(2.0\pm0.4)~{\rm GeV}^2\;,\hspace{1.0cm} G^K_{+,3/2}=-(1.0\pm0.45)~{\rm
GeV}\nonumber\\ &&K_{-,1/2}=-(1.2\pm0.20)~{\rm GeV}^2\;,
\hspace{0.8cm}G^K_{-,1/2}=-(1.6\pm0.6)~{\rm GeV} \,,\nonumber\\&&
\Sigma_{+,3/2}=(0.020\pm0.003)~{\rm GeV}^2\;,\hspace{0.63cm}
G^\Sigma_{+,3/2}=(0.013\pm0.007)~{\rm GeV}\;,\nonumber\\&&
\Sigma_{-,1/2}=(0.23\pm0.07)~{\rm GeV}^2\;,\hspace{1.0cm}
G^\Sigma_{-,1/2}=(0.042\pm0.034\pm0.053)~{\rm GeV} \,. \label{GK-numerical}
\end{eqnarray}

Let us evaluates numerically the sum rule for $\eta_{\rm ke}(y)$ and $\eta_{\rm
ke}^b(y)$ at first.  The continuum thresholds $\omega_{c1}$ and $\omega_{c2}$ in
(\ref{2-point1}) and (\ref{2-point2}) are determined by requiring stability of these
sum rules. One finds that $1.7~{\rm GeV}<\omega_{c1}<2.2~{\rm GeV}$ and $2.7~{\rm
GeV}<\omega_{c2}<3.2~{\rm GeV}$ \cite{neubert,huang}. Imposing usual criterion on
the ratio of contribution of the higher-order power corrections and that of the
continuum, we find that for the central values of the condensates and HQET
parameters given in (\ref{cond}) and (\ref{GK-numerical}), if the threshold
parameter $\omega_c$  lies in the range $1.9<\omega_c<2.5$ GeV, there is an
acceptable ``stability window'' $T=0.8-1.0$ GeV in which the calculation results do
not change appreciably. This window overlaps largely with those of the sum rules
(\ref{2-point1}) and (\ref{2-point2}). Therefore, our procedure of calculation is
justified. For estimating the errors induced by the uncertainties of parameters for
the condensates and HQET (\ref{cond}) and (\ref{GK-numerical}) we take the maxima
deviations from the central values of the condensates and HQET parameters and find
that for the existence of stability windows in the two extreme cases the continuum
thresholds shift to the range $2.5<\omega_c<2.9$ and $1.5<\omega_c<1.9$ GeV, respectively.
The corresponding windows for Borel parameters are $1.0<T<1.2$ and $0.6<T<0.8$ GeV, 
respectively. These are still compatible with the stability windows for the sum rules 
(\ref{2-point1}) and (\ref{2-point2}).

The numerical results of the form factors $\eta_{\rm ke}(y)$ and $\eta_{\rm
ke}^b(y)$ are shown in Fig. 2, where the curves refer to various choices for the
continuum thresholds and to the central values of the condensates and HQET parameters.

The numerical analysis shows that $\eta_{\rm ke}(y)$ is a slowly varying function in
the allowed kinematic range for $B\to D_1\ell\bar\nu$ and $B\to D_2^*\ell\bar\nu$
decays. The resulting curve for $\eta_{\rm ke}(y)$ may be well parameterized by the
linear approximation
\begin{eqnarray}\label{value1}
\eta_{\rm ke}(y)&=&\eta_{\rm ke}(1)\;(1-\rho_{\eta}^2(y-1))\;,
\hspace{0.1cm} \eta_{\rm ke}(1)=0.38\pm 0.17 \hspace{0.1cm}{\rm
GeV}\;, \hspace{0.1cm} \rho_{\eta}^2=0.8\pm 0.1 \;,\nonumber\\
\eta_{\rm ke}^b(y)&=&\eta^b_{\rm
ke}(1)\;(1-\rho_{\eta^b}^2(y-1))\;, \hspace{0.1cm} \eta_{\rm
ke}^b(1)=0.48\pm 0.21 \hspace{0.1cm} {\rm GeV}\;, \hspace{0.1cm}
\rho_{\eta^b}^2=1.0\pm 0.1 \;.
\end{eqnarray}

The final sum rules for $\eta_i(y)$ can be obtained by substituting Eq.
(\ref{sumrule-tau}) into Eq. (\ref{sum-rule2}) and (\ref{sum-rule4}). The numerical
evaluation for these sum rules proceeds along the same lines as that for $\eta_{\rm
ke}(y)$. Note that we have not included the perturbative term of order $\alpha_s$,
which is the leading perturbative contribution for $\eta_i (y)$. The sum rules for
$\eta_i(y)$ are not quantitatively reliable. Nevertheless, they are of correct order
of magnitude. The values of the form factors $\eta_1(y)$ and $\eta_3(y)$ at zero
recoil as functions of the Borel parameter are shown in Fig. 3, for three different
values of the continuum threshold $\omega_c$. The numerical results for $\eta_1(y)$
and $\eta_3(y)$ at zero recoil in the working regions read
\begin{eqnarray}\label{value2}
\eta_{1}(1)=-0.95\times 10^{-2}\;,\hspace{1cm} \eta_{3}(1)=3.5\times 10^{-2}\;
\end{eqnarray}
This result is in agreement with the expectation based on HQET that the
spin-symmetry violating corrections described by $\eta_i(y)$ are negligibly small.

Using the forms of linear approximations for $\eta_{\rm ke}(y)$ together with
$\tau(y)$ and $\tau_{1,2}(y)$ given in Ref. \cite{h-dai,hldai}
\begin{eqnarray}\label{tau}
\tau(y)=0.74(1-0.9(y-1))\;, \hspace{0.1cm} \tau_1(y)=-0.4(1-1.4(y-1))\;, \hspace{0.1
cm} \tau_2=0.28(1-0.5(y-1))
\end{eqnarray}
and neglecting the contribution of chromomagnetic correction, we can calculate the
total semileptonic rates and decay branching ratios by integrating Eqs.
(\ref{drate1}) and (\ref{drate2}). We use the  physical masses, $m_B=5.279$,
$m_{D_1}=2.422$ and $m_{D_2^\ast}=2.459$ \cite{PDG}, for $B$, $D_1$ and $D^*_2$
mesons. The maximal values of $y$ in the present case are
$y^{D_1}_{max}=(1+r_1^2)/2r_1\approx 1.32$ and $y^{D^*_2}_{max}=(1+r_2^2)/2
r_2\approx 1.31$. The quark masses are taken to be $m_b=4.8$ GeV, $m_c=1.5$ GeV. In
Table \ref{tab:branch} we present our results for decay rates both in the infinitely
heavy quark limit and taking account of the first order $1/m_Q$ corrections as well
as their ratio
\begin{eqnarray}
R_{\infty}=\frac{{\rm Br}(B\to D^{**}\ell\nu)_{{\rm with}\, 1/m_Q}}{{\rm Br}(B\to
D^{**}\ell\nu)_{m_Q\to\infty}}\;.
\end{eqnarray}

From Table \ref{tab:branch} we see that the $B\to D_1\ell\bar\nu$ decay rate
receives large $1/m_Q$ contributions and gets a sharp increase, while the $B\to
D^*_2\ell\bar\nu$ decay rate is only moderately increased by subleading $1/m_Q$
corrections. The reason for this is as following. From Eqs. (\ref{drate1}),
(\ref{drate2}) and (\ref{fkexp}) we see that $(y-r_1)^2 f_{V_1}^2$ term dominates
the differential width for decay to $D_1$ near $y=1$. $f_{V_1}$ vanishes at the
leading order and receives non-vanishing contributions from first order heavy quark
mass corrections:
\begin{eqnarray}\label
{fv1}\sqrt{6} f_{V_1}(1)=-8\varepsilon_c(\bar\Lambda'-\bar\Lambda)\tau(1).
\end{eqnarray}
Since the allowed kinematic ranges for $B\to D_1\ell\bar\nu$ is fairly small, the
contribution to the decay rate of the $1/m_Q$ corrections is substantially
increased. On the other hand, the matrix elements (\ref{d2-a}) and (\ref{d2-v}) of
the $B\to D^*_2\ell\bar\nu$ decay vanish at zero recoil without using the heavy mass
limit. The term $(y-r_2)^2 k_{A_1}^2$ dominates the $B\to D^*_2\ell\bar\nu$ decay
rate, but $k_{A_1}$ does not vanish at the leading order recoil. Therefore, this
process is not much affected by next-to-leading corrections. Note that although the
correction to the rate for decay to $D_1$ is very large it comes mainly from the
effect of the different masses of $B$ and $D_1$. The values of $\eta_{ke}$ and
$\eta_ {ke}^b$ in (23) are of the order $\Lambda_{QCD}$ and perfectly normal.

In Table~\ref{tab:branch} the available experimental data for
semileptonic $B$ decay to excited $D^{**}$ mesons are presented.
As for the $B\to D_2^*\ell\bar\nu$ branching ratio  there are only
upper limits from these experimental groups except the data from
OPAL. In comparison with the experimental data our result for the
branching ratio of the $B\to D_1\ell\bar\nu$ decay with the
inclusion of $1/m_Q$ corrections is larger than the CLEO and ALEPH
measurements but is consistent with OPAL and DELPHI data. On the
other hand, our branching ratio for the $B\to D^*_2\ell\bar\nu$
decay  disagrees with the ALEPH data but is consistent with
results from other groups.

%%%%%%%%%%%%%%%%%%%%%%%%%%%%%%%%%%%%%%%%%%%%%%%%%%%%%%%%%%%%%%%%%%%%%%%%%%%
\section{Conclusion}\label{conc}
In this work we have presented the investigation for semileptonic
$B$ decays into excited charmed mesons. Within the framework of
HQET we have applied the QCD sum rules to calculate the universal
Isgur-Wise functions up to the subleading order of the heavy quark
expansion. The Isgur-Wise functions $\eta_{\rm ke}$ and $\eta_{\rm
ke}^b$ related to the insertions of kinetic energy operators of
the HQET Lagrangian are found of normal values of the order
$\Lambda_{QCD}$, while the form factors $\eta_i$, parameterizing
the time-ordered products of the chromomagnetic operator in the
HQET Lagrangian with the leading order currents, are negligibly
small.These results are in agreement with the HQET-based
expectations.

We have computed, for the decays $B\to D_1\ell\bar\nu$ and $B\to
D_2^*\ell\bar\nu$, the differential decay widths and the branching
ratios with the inclusion of the order of $1/m_Q$ corrections. Our
numerical results show that the  first order $1/m_Q$ correction is
not large for the decay rate of $B\to D_2^*\ell\bar\nu$ process,
but is very large for the $B\to D_1\ell\bar\nu$ process. We have
explained the reason for this result.
%%%%%%%%%%%%%%%%%%%%%%%%%%%%%%%%%%%%%%

\acknowledgments M-Q H would like to thank Chun Liu for helpful discussions. M-Q H's
work is supported in part by the National Natural Science Foundation of China under
Contract No. 19975068. Y-B D's work is supported by National Natural Science
Foundation of China. \vspace{0.5cm}

\appendix
\section{}\label{appendix}
We list here the  lorentz structures used in the paper.
\begin{eqnarray}
 {\cal L}^{\mu\nu}_{V}&=&\frac{1}{\sqrt{6}}\left[(y^2-1)g_t^{\mu\nu}+(3v^\mu+
 (2-y)v'^\mu)v_t^\nu\right]\;,\\
 {\cal L}^{\mu\nu}_{A}&=&i\frac{1}{\sqrt{6}} (1+y)\epsilon ^{\mu\nu'\alpha\beta}
 g_t^{\nu\nu'}v_\alpha v'_\beta\;,\\
 {\cal L}^{\mu\alpha\beta}_{V}&=&-\frac{i}{2}\epsilon^{\mu\nu\sigma\rho}\left(
 v_{t}^{\alpha}g_{t}^{\beta\nu}+g_{t}^{\alpha\nu}v_{t}^{\beta}\right)v_\sigma v'_\rho\;,\\
{\cal L}^{\mu\alpha\beta}_{A}&=&-\frac{1}{2}(1+y)\left(v_t^\alpha
g_t^{\mu\beta}+v_t^\beta g_t^{\mu\alpha}-\frac{2}{3}v_t^\mu
g_t^{\alpha\beta}\right)\nonumber\\&&\quad\mbox{} +v^{\prime\mu}\left(v_t^\alpha
v_t^\beta-\frac{1}{3}(1-y^2)g_t^{\alpha\beta}\right)\;,\\{\cal L}^{\mu\nu}_
{V\xi}&=&-\frac{1}{\sqrt{6}}(v^\mu+v'^\mu)v_t^\nu \;,\\ {\cal L}^{\mu\alpha\beta}_
{A\eta_2}&=&(-v^\mu+v'^\mu)\left(2v_t^\alpha
v_t^\beta-\frac{2}{3}(1-y^2)g_t^{\alpha\beta}\right)\;, \label{lorentz1}
\end{eqnarray}
where $g_t^{\alpha\beta}=g^{\alpha\beta}-v^{\prime\alpha} v^{\prime\beta}$ and
$v_t^\alpha=v^\alpha-yv'^\alpha$.

\newpage
{\bf Figure Captions} \vspace{2ex}
\begin{center}
\begin{minipage}{120mm}
{\sf Fig. 1.} \small{Feynman diagrams contributing to the sum rules for the
                    Isgur-Wise form factor in the coordinate gauge. The gray
                    square corresponds to the insertion of the kinetic energy operator
                    at ${\cal{O}}(1/m_Q)$ in the HQET Lagrangian.  }
\end{minipage}\end{center}

\begin{center}
\begin{minipage}{120mm}
%\begin{minipage}{120mm}
{\sf Fig. 2.} \small{ Results of the numerical evaluation for the sum rules:
Isgur-Wise form factors $\eta_{\rm ke}(y)$ and $\eta_{\rm ke}^b(y)$ with $T=0.9$
GeV.}
\end{minipage}\end{center}
%\end{document}

\begin{center}
\begin{minipage}{120mm}
%\begin{minipage}{120mm}
{\sf Fig. 3.} \small{Dependence of $\eta_1(1)$ and $\eta_2(1)$ on the Borel
parameter $T$ for different values of the continuum threshold $\omega_c$.}
\end{minipage}\end{center}

%%%%%%%%%%%%%%%%%%%%%%%%%%%%%%%%table I here%%%%%%%%%%%%%%%%%%%
\begin{table}\centering\begin{minipage}{13 truecm}
\caption{Decay rates $\Gamma$ (in $10^{-15}$ GeV) for $|V_{cb}|=0.04$ and branching
ratios BR (in \% and taking $\tau_B=1.6$ps) for $B\to D^{**}\ell\nu$ decays in the
infinitely heavy quark mass limit and taking account of first order $1/m_Q$
corrections.} \label{tab:branch}

%\begin{minipage}{13 truecm}
\begin{tabular}{cccc}
%\hline
%\multicolumn{2}{c}{}
&&$B\to D_1e\nu$&$B\to D^*_2e\nu$\\\hline\hline
%%%%%%%%%%%%%%%%%%%%%%%%%%%%%%%%%%%%%%%%%%%%%%%%%%%%%%%%%%%%%
$m_Q\to\infty$&$\Gamma$&1.4&2.1\\ &Br&0.34&0.52\\\hline
%%%%%%%%%%%%%%%%%%%%%%%%%%%%%%%%%%%%%%%%%%%%%%%%%%%%%%%%%%%%%
With $1/m_Q$&$\Gamma$&5.3&2.4\\ &Br&1.3&0.59\\\hline
%%%%%%%%%%%%%%%%%%%%%%%%%%%%%%%%%%%%%%%%%%%%%%%%%%%%%%%%%%%%%
%\multicolumn{2}{c}{$R_\infty$}
$R_\infty$&&3.78&1.15\\\hline
%%%%%%%%%%%%%%%%%%%%%%%%%%%%%%%%%%%%%%%%%%%%%%%%%%%%%%%%%%%%%
Experiment&Br (CLEO) \cite{cleo}~~~ &$0.56\pm 0.13\pm0.08\pm0.04$&$<0.8$\\ &Br (ALEPH)
\cite{aleph}~ &$0.74\pm0.16$&$<0.2$\\ &Br (OPAL)
\cite{opal}~~~&$2.0\pm0.6\pm0.5$&$0.88\pm0.35\pm0.17$\\&Br (DELPHI)
\cite{delphi}&$1.5\pm0.55$&$<6.25$\\
\end{tabular}\end{minipage}
\end{table}

%%%%%%%%%%%%%%%%%%%%%%%%%%%%%%%%%%%%%%%%%%%%%%%%%%%%%%%%%
%%%%%%%%%%%%%%%%Fig. 1%%%%%%%%%%%%%%%%%%55
\begin{figure}[htbp]   % produce figure here
\begin{center}
\setlength{\unitlength}{1truecm}
\begin{picture}(6.8,6.8)%(<right,>top)
\put(-7.0,-18) {\includegraphics{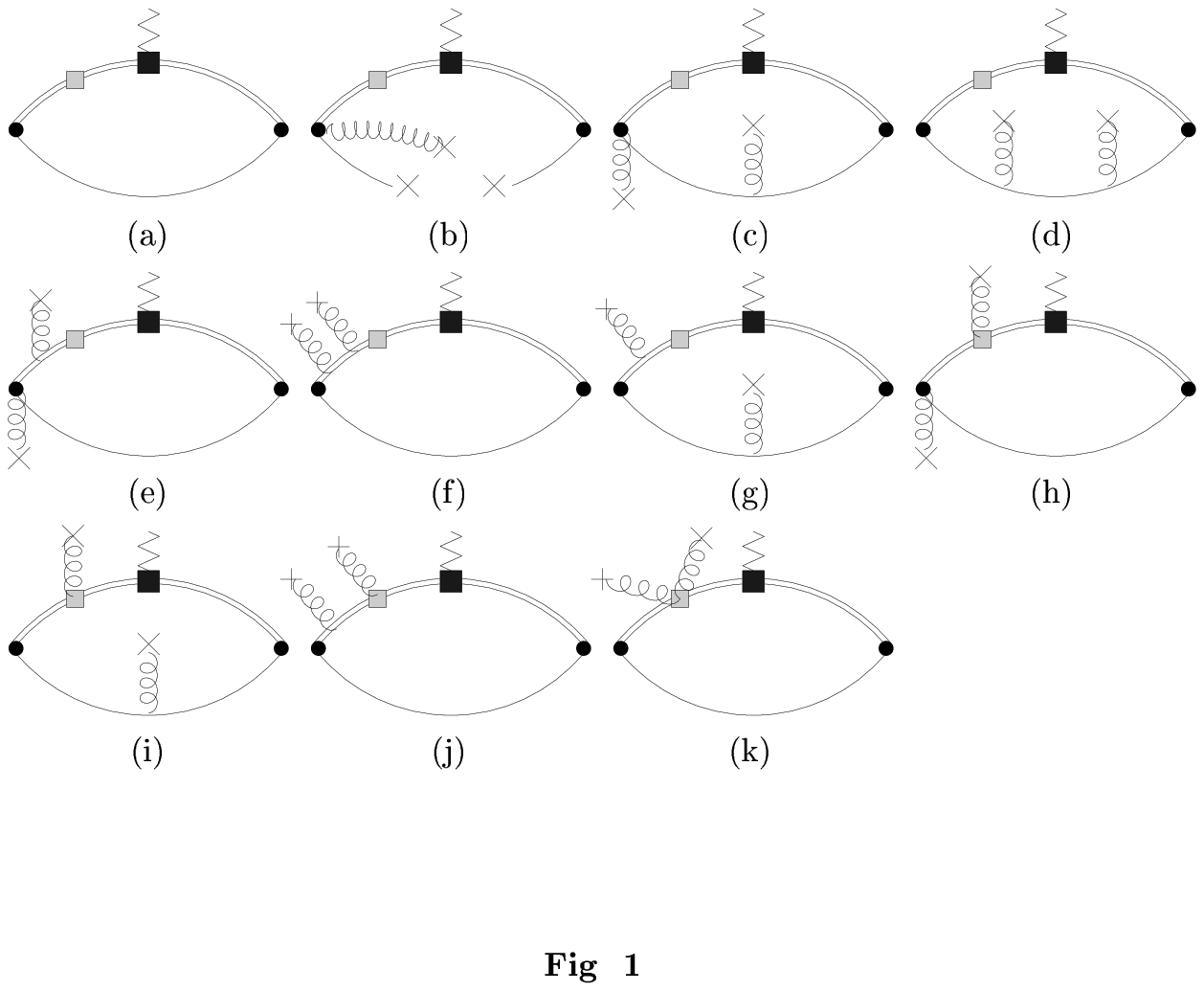}}
\end{picture}
\end{center}
\vskip 2.0cm
%\fcaption{xx}
\protect\label{Fig.1}
\end{figure}
%%%%%%%%%%%%%%%%Fig. 2%%%%%%%%%%%%%%%%%%%
\newpage
\begin{figure}[htbp]   % produce figure here
\begin{center}
\setlength{\unitlength}{1truecm}
\begin{picture}(6.8,6.8)%(<right,>top)
\put(-6.0,-14) {\includegraphics{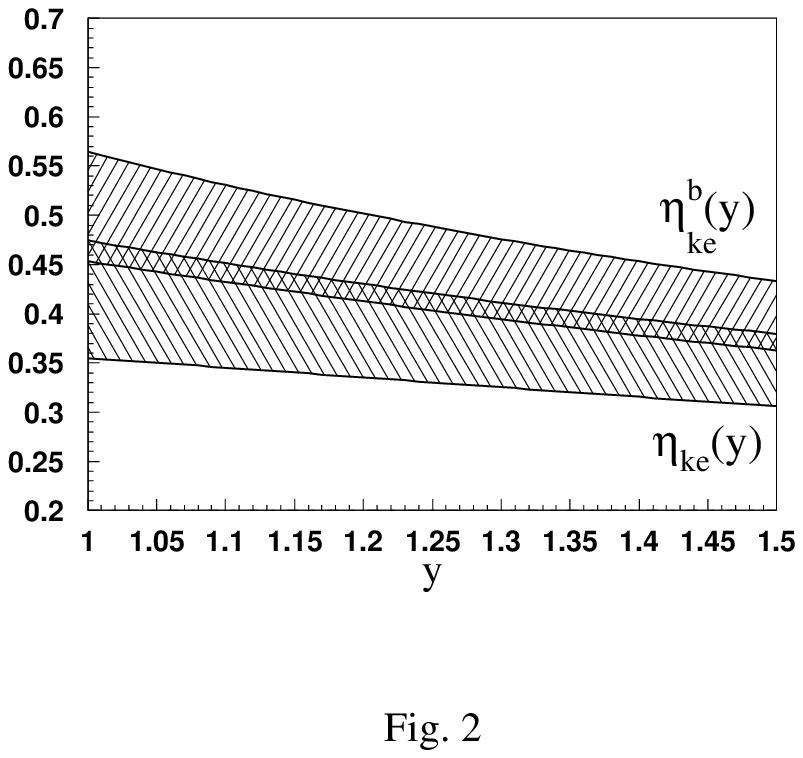}}
\end{picture}
\end{center}
\vskip 2.0cm
%\fcaption{xx}
\protect\label{Fig.2}
\end{figure}
%%%%%%%%%%%%%%%%Fig. 3%%%%%%%%%%%%%%%%%%%
%\newpage
\begin{figure}[htbp]   % produce figure here
\begin{center}
\setlength{\unitlength}{1truecm}
\begin{picture}(6.8,6.8)%(<right,>top)
\put(-6.0,-14) {\includegraphics{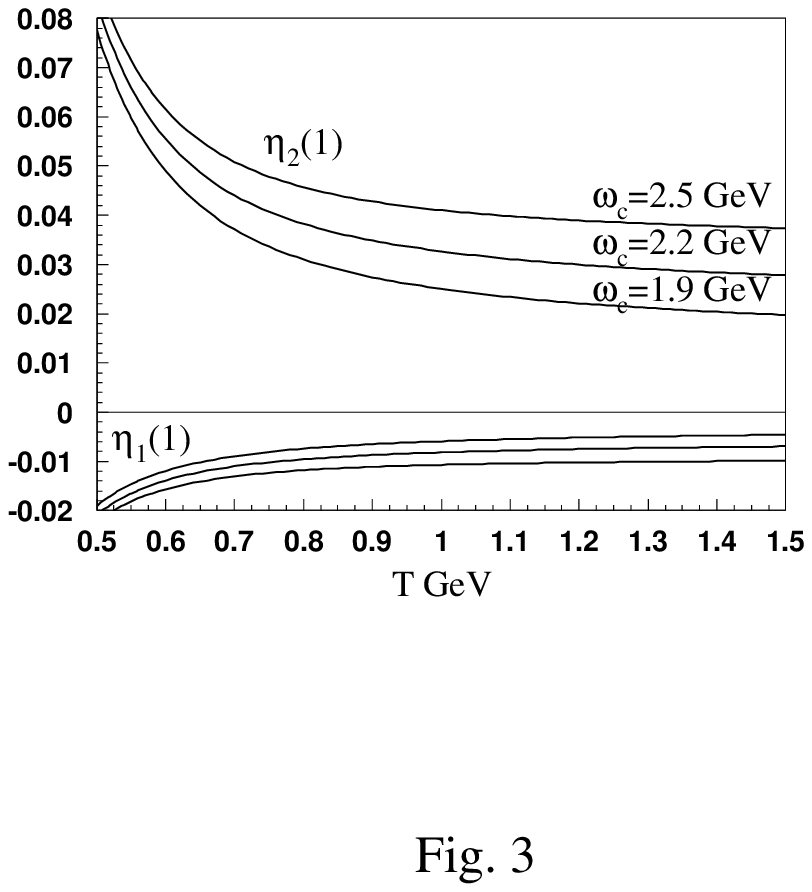}}
\end{picture}
\end{center}
\vskip 2.0cm
%\fcaption{xx}
\protect\label{Fig.3}
\end{figure}
\end{document}